\documentclass[12pt,preprint]{iopart}
\usepackage{graphicx,latexsym}
\usepackage{dcolumn}
\usepackage{amssymb,bm}
\usepackage{xcolor}
\usepackage{mathrsfs}
\usepackage[T1]{fontenc} 
\usepackage[utf8]{inputenc}
\usepackage{amsgen}
\usepackage{amsfonts}
\usepackage{amsbsy}
\usepackage{epsfig}
\usepackage{lineno}
\usepackage{multicol}
\DeclareUnicodeCharacter{0301}{\"{e}}
\newcommand{\beq}{\begin{equation}}
\newcommand{\eeq}{\end{equation}}
\begin{document}

\title{Dynamics of Rapidly Rotating Bose-Einstein Quantum Droplets}

\author{Szu-Cheng Cheng$^1$, Yu-Wen Wang$^2$, and Wen-Hsuan Kuan$^2$}
\address{$^1$Department of Physics, Chinese Culture University, Yang-Ming-Shan, Taipei 11114, Taiwan.}
\address{$^2$Department of Applied Physics and Chemistry, University of Taipei, Taipei 10048, Taiwan.}
\ead{wenhsuan.kuan@gmail.com}

\date{\today}

\begin{abstract}
   The study presents a detailed analysis of the stationary properties and dynamics of an anharmonically trapped, rapidly rotating Bose-Einstein liquid droplet. We investigate the effects of the particle number, confining potential, and rotation speed on the formation of the energetically favored bead, multiple quantized vortex, off-center vortex, and center-of-mass states. The multi-periodic trajectories and breathing provide evidence of the collective excitations of the surface mode in the vortex states. Observation of the self-trapped phenomenon and tendency toward the lowest Landau level in rapid rotation regimes coincides well with the quantum-Hall limit. Modifying the topological charges and destroying the potential flow of the vortex state can occur if an external disturbance is imposed upon the quantum droplet.
\end{abstract}

\maketitle

\section{Introduction}
%
Dilute atomic gases have been created at extremely low temperatures using laser cooling and trapping techniques. According to the grand canonical ensemble theory, the critical temperature for specifying the quantum phase transition can be defined by the phase-space filling at zero chemical potential, and the macroscopic collection of cold atoms with a nonzero off-diagonal long-range order forms Bose-Einstein condensates (BECs) \cite{Anderson1995, Bradley1995, Davis1995}. With the mean field (MF) approximation, the dynamics of BECs with weak interatomic interactions can be adequately described by applying the Gross-Pitaevskii equation (GPE). Near the Feshbach resonance, where elastic scattering can be dramatically altered by an external field \cite{Jin2004, Ketterle2004, Hulet2005}, a quasibound molecule can tunnel across a potential energy barrier and resonantly couple with the free states of colliding atoms. Therefore, the possibility of tuning the magnitude and sign of the scattering length using external magnetic fields provides new perspectives for manipulating BECs \cite{Anderson1995, Eagles1969, Julienne2006}.

Concerning two-body interactions in a monatomic ensemble, an equilibrium between the attractive interatomic forces and short-range repulsion due to van der Waals forces forms a liquid instead of a gas during the cooling process. Owing to their high density and incompressibility, attempts by usual liquids to enter the quantum regime are prohibited.      
However, with the inclusion of the Lee-Huang-Yang (LHY) correction to the ground state energy of a homogeneous weakly repulsive Bose gas \cite{LHY1957}, a new type of quantum liquid that emerges in ultracold and extremely dilute atomic systems and violates the van der Waals model, has recently been observed in two-component Bose mixtures \cite{Science359-301-2018, PRL120-13501-2018, Semeghini2018} and single-component dipolar condensates\cite{Barbut2016, Schmitt2016, Barbut2018, Chomaz2016}.  
Above the particle number threshold, the gas-to-liquid phase transition occurs when instabilities arising from attractive mean-field interaction $\propto n^2$ are compensated for by high-order quantum many-body effects $\propto n^{5/2}$ \cite{PRL115-155302-2015}. 
By tuning the interatomic interactions, the realization of dilute and weakly interacting self-bound liquid droplets provides direct evidence of the beyond MF effects \cite{FP16-32201-2021}.

As a model for quantum liquids, Bose superfluids exhibit exceptional coherence, as demonstrated by the calculation of the one-body density matrix and correlation functions between any two particles. For spinless atomic gases, the phase gradient of the condensate order parameter defines the superfluid velocity, which is irrotational, unless there is a phase singularity such that the Onsager-Feynman quantization condition is satisfied. 
Through laser-stirred phase imprinting, the generation of vortex cores \cite{Matthews1999, Madison2000}, vortex rings \cite{Anderson2001, Dutton2001}, and even vortex lattices \cite{Madison2001, Abo2001} in rapidly rotating dilute-gas BECs has been experimentally realized. \textcolor{black}{Meanwhile, it was first predicted by Kasamatsu et al. that a giant vortex can be formed by absorbing all phase singularities into the low density hole region \cite{Kasamatsu2002}.} In tightly trapped BECs, the formation of multiple quantized vortices and long-lived vortex aggregations has also been observed \cite{Lundh2002, Engels2003}, demonstrating the crucial influence of Coriolis forces in rotating systems.

In contrast to gaseous BECs, theoretical analysis has demonstrated that quantum droplets (QDs) of singly charged vortex states are unstable in single-component dipolar condensates \cite{Cidrim2018, Malomed2019}. However, it was reported that by analyzing the linearized Bogoliubov-de Gennes equations, stable solutions can theoretically be found for 2D QDs with hidden and explicit vorticity \cite{Li2018} and for 3D binary condensates with contact and LHY-amended interactions \cite{Kartashov2018}. \textcolor{black}{Various phases with exotic singularities in rotating droplets confined in harmonic and anharmonic traps are also identified \cite{Nikolaou2023, Nikolaou2024}.} It has also been experimentally shown that the application of optical lattices helps stabilize zero-vorticity and vortical solitons \cite{Yang2003} whenever the Vakhitov-Kolokolov criterion is violated in vortex lattice QDs \cite{Zheng2021}. The formation of vortex lattices in a Gaussian-like QD \cite{Thiruvalluvar2019} and the quantum Hall limit \cite{Gu2023} have also been reported. Regarding these quantum liquids, the particle number of the systems with relevant finite-size effects is critical for the formation of low-energy modes. 


Interestingly, an electron system confined to two dimensions and subjected to a perpendicular magnetic field presents a harmonic oscillation problem. In this system, the electron kinetic energy is quantized and the discrete energy levels are termed Landau levels. The lowest Landau level (LLL) occupation may occur under strong magnetic field regimes. In contrast, the presence of interactions and disorders may stimulate excitation to the higher Landau level states, resulting in Landau level mixing (LLM), which is prominent for a strictly 2D system because the primary contribution due to the short-range interaction is more effective with decreasing thickness. From this point of view, the integral quantum Hall effect that exhibits quantized Hall resistance plateaus and Arrhenius behavior in the longitudinal resistance \cite{Klitzing1980, Klitzing1986} is a direct consequence of Landau level formation. For fractional quantum Hall systems, the introduction of Laughlin's theory and composite fermion principle provides crucial clues for exploring novel phenomena such as skyrmions and fractional statistics \cite{Stormer1999, Tsui1982, Tsui1999}.

By sharing many similarities, the application of the same mathematical formalism to systems of charged particles in a strong magnetic field and rotating neutral atoms has successfully predicted the bosonic Laughlin state \cite{Cooper1999, Laughlin1983}. Some novel phenomena, such as the emergence of incompressible states, bosonic analogs of the Jain sequence \cite{Jain2007}, and Abelian and non-Abelian states \cite{Moore1991, Read1999}, have been extensively explored in the past decade. In addition, the formation of exotic states in rotating spinor bosons has been studied \cite{Reijnders2002}. In recent years, the realization of geometric squeezing into the LLL of BECs and interaction-driven spontaneous crystallization has offered a new route towards strongly correlated fluids and bosonic quantum Hall states \cite{Fletcher2022, Mukherjee2022}.
While the discovery of quantum Hall effects has revealed remarkable macroscopic quantum phenomena related to topological investigations of vortices, the systems of rotating BECs with multiple degrees of freedom of tunable parameters are versatile and can be promising for topological quantum computations \cite{Nayak2008}.
%

This study aims to theoretically investigate the ground state properties and dynamics of a 2D rapidly rotating QD confined in a symmetric anharmonic trap. 
In the presence of a rotation field, radial confinement is reduced by the centrifugal potential, and the Coriolis force experienced by the droplets in the rotating frame is equivalent to the Lorentz force on a charged particle. 
As the analogy between quantum Hall effects and rapidly rotating BECs in dilute gases has been precisely recognized, it will be interesting to see whether artificial Lorentz forces can also be engineered for a new type of quantum liquid such that the generation of exotic phases carrying nonzero topological charges can be observed, the route for the quantum phase transition can be depicted, and the singularity and stability of the quantum states can be managed in this system. 
These explorations would make it possible to simulate quantum Hall-type effects in a controlled manner with low-dimensional non-uniform quantum liquids. Because the combined impact of interactions and quantum statistics eventually determines the features of the many-body ground state, such rotating systems allow for the study of artificial orbital magnetism in quantum liquids.

The remainder of this paper is organized as follows. In Sec. II, we first introduce the criterion to form 2D QDs and then present the analytical solutions of the LHY-amended GPE for an anharmonically trapped rotating QD utilizing the variational method with the inclusion of the LLM effect. In Section II.A, we analyze the stationary properties, including density profiles and phase distributions. A brief comparison of the rotating QD and optical vortex is presented in Section II.B to better interpret the underlying physics of phase distributions. 
The effects of nonlinear interaction and rotation speed on the ground state configuration are demonstrated using the phase diagram in Section II.C. A formulation for the maximum circulation restriction was proposed based on the LLL approximation to avoid collapse from three-body collisions and numerical divergence. 
In Sec. III, we further investigated the dynamics of the rotating QD. By solving the Euler-Lagrange equations, we analyzed the periodicity and stability of the QD under linear perturbation in Section III.A. and study superfluids with probability current density distributions of a long-term evolutionary vortex state in Section III.B. Finally, we summarize our work in Sec. IV.

\section{Theoretical Model for Rotating Quantum Droplets}
At extremely low temperatures, the mutual interactions between ultra-dilute gaseous atoms are well described in terms of the hard-core model and s-wave scattering length. 
However, the situation becomes more complicated for low-dimensional liquid systems, even though the 3D s-wave scattering lengths $a^{3D}$ and $a^{3D}_{\uparrow\downarrow}$ adopted in Refs.~\cite{Semeghini2018} and \cite{PRL126-230404-2021} can be used to suitably describe the liquid-like droplet regimes. 
To form a 2D quantum droplet in a system of dilute Bose-Bose mixtures with densities $n_{\uparrow}$ and $n_{\downarrow}$, a weakly attractive interaction for the interspecies and a weakly repulsive interaction for the intraspecies are required. Using the scattering \textit{t} matrix \cite{Popov1972}, the weakly interacting regime beyond the MF approximation can be approached theoretically \cite{Petrov2016, PRA98-051604R-2018}.  
For the symmetric case $a_{\uparrow \uparrow}=a_{\downarrow \downarrow}=a$ and $n_{\uparrow}=n_{\downarrow}=n$, the energy density of a uniform liquid under the assumption of a macroscopic condensate population can be derived analytically. 

Referring to Ref.~\cite{Petrov2016}, the 2D scattering lengths $a$ and $a_{\uparrow\downarrow}$ would take the form of $a = l_z \exp[-\sqrt{\pi/2}\,l_z/a^{3D}]$ and $a_{\uparrow\downarrow} = l_z \exp[-\sqrt{\pi/2}\,l_z/a^{3D}_{\uparrow\downarrow}]$, respectively, where $l_z = \sqrt{\hbar/2M\omega_z}$ is the oscillator length in the strong confinement direction. 
In this study, the simulation of a symmetric mixture of a quasi-2D $^{39}$K oblate droplet with $a^{3D}_{\uparrow\downarrow} = -53\, a_0$, and $a^{3D} = 50\, a_0$ \cite{Science359-301-2018} was performed. For a typical experimental parameter $\omega_z = 2\pi\times 400$\, Hz, it was found that the mixtures could be transformed from stable gases to liquids by introducing confinement. The system of dilute liquid is considered because the inequality $1/\ln(a_{\uparrow\downarrow}/a)=1.8\times 10^{-3} \ll 1$ is satisfied.    
For the confined QDs, the typical length scale on which the wavefunction changes is in order of the healing length $\xi \sim \hbar/\sqrt{M|\mu|}$ that represents the vortex core size, where $\mu \sim -n^2\hbar^2/M\ln^2(a_{\uparrow \downarrow}/a)$ is the chemical potential. For liquid droplets, the surface tension is crucial to the effects of finite size on the energy and surface mode spectra. 


%
The rotating QDs were analyzed in the non-inertial reference frame based on the unitary transformation $H_r = H_l - \vec{\Omega}\cdot \vec{L}$, where $H_r$ and $H_l$ are the rotating and lab frame Hamiltonians, respectively; $\vec{\Omega}$ is the angular velocity; and $\vec{L}$ is the orbital angular momentum (OAM) of the object.
In this manner, the time-dependent GPE with the inclusion of high-order quantum fluctuations on condensed cold atoms for a symmetric anharmonically-trapped rotating QD can be written as 
%
%
%
\begin{equation}
\hspace{-3cm}
i \hbar \frac{\partial \Psi}{\partial \tau} = \frac{1}{2M}(-i\hbar\vec{\nabla} - M\Omega\hat{z}\times\vec{r})^2 \Psi + \left[\frac{1}{2}M(\omega_0^2-\Omega^2) r^2 + \frac{1}{4}\lambda' r^4\right] \Psi + \sigma' |\Psi|^2 \ln(|\Psi|^2/n_0)\,\Psi,  
\label{Eq-2}
\end{equation}
in which $\omega_0$ is the trapping frequency of the harmonic potential and $\lambda'$ is the strength parameter of the quartic trap. \textcolor{black}{Under local density approximation, the last term specifies the LHY-amended nonlinear interaction, where} $\sigma' = {8 \pi \hbar^2}/{M\ln ^2\left(a_{\uparrow \downarrow} / a\right)}$ and $n_0$ is the equilibrium density of the component. The first term on the right side of Eq.~(\ref{Eq-2}) precisely shows that the rotating droplet moves as a charge $q$ in the $xy$ plane subjected to a synthesized magnetic field $B\hat{z}$ with a vector potential $(q/c)\vec{A} = M\Omega\hat{z}\times \vec{r}$, which associates the cyclotron frequency $\omega_c = Bq/Mc$ and defines the magnetic length $l_c = \sqrt{\hbar c/qB}$. 
The solutions of non-interacting and harmonically-trapped systems provide eigenenergies 
\beq
E_{n_l,m} = (2 n_l + |m| + 1)\hbar\omega_0 - m\hbar\Omega
\eeq
and the corresponding eigenfunctions   
\beq
\hspace{-2.5cm}
\chi_{n_l,m}(\vec{r}) = \frac{\sqrt{b}}{\sqrt{2\pi l_c^2}}\sqrt{\frac{n_l !}{2^{|m|}(|m|+n_l)!}}\,\left(\frac{\sqrt{b}r}{l_c}\right)^{|m|} e^{-br^2/4\,l_c^2}\,e^{-i m\phi} L_{n_l}^{|m|}\left(\frac{br^2}{2\,l_c^2}\right), \label{Eq-LL}
\eeq
where $b = (1+ 4(\omega_0^2-\Omega^2)/\omega_c^2)^{1/2}$, $L_{\nu}^k(t)$ is the associated Laguerre polynomial, integers \textcolor{black}{$n_l = 0 , 1, 2, \ldots$} denote the Landau level index; and \textcolor{black}{$m = 0, \pm 1, \pm 2, \ldots$} corresponds to the degenerate states within a Landau level.


The ground-state properties of the rotating QDs were examined using the semiclassical variational method. For simplicity, we rewrite Eq.~(\ref{Eq-2}) in a dimensionless form by taking $\Psi = \sqrt{n_0}\Psi$, rescaling the energy, length, time, and quartic parameter by $\hbar\omega = \sigma' n_0$, $l = \hbar/\sqrt{2M\sigma'n_0}$, $\tau = t/\omega$, and $\lambda = \hbar/M^2\omega^3 \lambda'$, respectively, and denoting $\omega_0/\omega = \omega_0$ and $\Omega/\omega = \Omega$. Hence, the energy functional $E[\Psi,\Psi^*]$ can be written as:
\begin{eqnarray}
\hspace{-2cm}
E[\Psi,\Psi^*] &&= \int d\vec{r}\, \left[\Psi^*\left(-i\hbar\vec{\nabla} - \frac{1}{2}\hat{z}\times\vec{r}\right)^2 \Psi + \left[\frac{1}{4}(\omega_0^2 - 1) r^2 + \frac{1}{16}\lambda r^4\right] |\Psi|^2 \right. \nonumber\\
&& + \left. \frac{1}{2}|\Psi|^4 \ln(|\Psi|^2/\sqrt{e}) - \Psi^*
(\Omega - 1)L_z \Psi \right].  
\end{eqnarray}
While the presence of attractive inter-particle interactions would likely mix different $(n_l, m)$ states, the LLL approach \textcolor{black}{corresponding to $n_l = 0$} would be insufficient to describe the rotating droplets, especially for strongly confined atoms, because the density would not be thinned out in the $xy$ plane, as observed in vortex lattices that bear repulsive inter-particle interactions. 
Inspired by the work of a 2D Wigner crystal in a strong magnetic field \cite{Maki1983}, we set the trial wavefunction 
\vspace{-0.25cm}  
\begin{equation}
\hspace{-3cm}
\Psi = C_m (\mathfrak{z} - \mathfrak{z}_0)^m \exp\left[ -\frac{(\vec{r}-\vec{R})^2}{4\rho^2} - \frac{i}{2}\hat{z}\cdot(\vec{r}\times \vec{R})\right] \exp\left[i \frac{\alpha}{4} (\vec{r}-\vec{R})^2 \right] \exp\left[i \frac{\vec{p}}{2}\cdot (\vec{r}-\vec{R}) \right], \label{Eq-wf0}
\end{equation} 
where $\mathfrak{z}$ and $\mathfrak{z}_0$ are the complex notations of the 2D position vectors, and $\rho$ sketches the width of the wavepacket and is representative of the influence of the LLM effect that should be addressed unless an extremely high magnetic field or a rapid rotation is reached, where the assumption of the LLL with $\rho = 1$ is feasible. \textcolor{black}{Owing to a Gaussian-like envelope function, Eq.~\ref{Eq-wf0} is suitable for describing systems with limited particle numbers. In contrast, a flat-top or a super-Gaussian function would be appropriate in the Thomas-Fermi limit.}  
Taking the polar coordinate representation, the azimuthal phase $\exp(i m\phi)$ with an integer number $m$ embedded in the power function $(\mathfrak{z}-\mathfrak{z}_0)^m$ introduces the quantized OAM, as well as the synthetic magnetic flux quanta carried by a rotating QD. 
In this ansatz, $\vec{r} = x \hat{x} + y\hat{y}$ is the position of the particle in the droplet and $\vec{R} = x_0 \hat{x} + y_0 \hat{y}$ denotes its center-of-mass position. The phase associated with $\vec{r}\times \vec{R}$ allows rigid-body rotation and ensures gauge invariance in a synthetic magnetic field. 
The parameters $\alpha$ and $\vec{p}$ in the last two parentheses denote the conjugate curvature coefficient and momentum, respectively, which represent the inherent MF expansion and relative repulsion of the wavepacket, respectively. This makes it possible to further investigate the variation in the density profiles and coherent properties during the time evolution of the QDs.
In the absence of dissipation due to three-body collisions, the particle number is conserved, and $C_m = \sqrt{N 2^{m+1}/\pi m!}/\rho^{m+1}$. 

For convenience, we mapped the system onto a 2D complex plane setting the complex coordinates $\mathfrak{z}_0 = \frac{1}{2} (x_0 + i y_0)$, $\mathfrak{z} = \frac{1}{2} (x + i y)$, and momentum $\mathfrak{p} = \frac{1}{2} (p_x + i p_y)$, such that $|\mathfrak{z}_0|^2 = R^2/4$, $|\mathfrak{z}|^2 = r^2/4$, and $|\mathfrak{p}|^2 = p^2/4$ build complex-real connections. By using the reconstructed wavefunction  
\begin{eqnarray} 
\hspace{-1cm}
\Psi = C_m (\mathfrak{z}-\mathfrak{z}_0)^m \exp\left[-\frac{1}{\rho^2}|\mathfrak{z}-\mathfrak{z}_0|^2\right] 
\exp\left[ \mathfrak{z}^* \mathfrak{z}_0 - \mathfrak{z} \mathfrak{z}_0^* \right]  
\exp\left[ i \alpha |\mathfrak{z}-\mathfrak{z}_0|^2 \right] &\nonumber\\
   \times \exp\left[ i \mathfrak{p}^* (\mathfrak{z}-\mathfrak{z}_0) + i\mathfrak{p} (\mathfrak{z}^* - \mathfrak{z}_0^*)\right], & \label{eq-wf}
\end{eqnarray}
the energy functional per particle is given by  
\begin{eqnarray}
\fl
 \mathscr{E}[\Psi,\Psi^*] = \mathscr{E}_{kin} + \mathscr{E}_{trap} + \mathscr{E}_{non} + \mathscr{E}_{rot} &\nonumber\\
\fl
 = \biggl\{\frac{(m+1)}{2}\left(\rho^2 + \frac{1}{\rho^2} + \alpha^2\rho^2 \right) - m + |\mathfrak{p}|^2 \biggr\} 
+ \biggl\{\left({\omega_0^2} - 1\right)\left(\frac{m+1}{2}\rho^2+|\mathfrak{z}_0|^2 \right) & \nonumber\\
\fl
 + \lambda\left[\frac{1}{4}(m+2)(m+1)\rho^4 + |\mathfrak{z}_0|^4 + 2(m+1)\rho^2|\mathfrak{z}_0|^2 \right]\biggr\}&  \nonumber\\
%
\fl
 + \biggl \{ \frac{N}{\pi(m!)^2 2^{2m+2}}\frac{1}{\rho^2}\biggl[ (2m)! \ln\left(\frac{|C_m|^2}{4^m\sqrt{e}}\right) - \frac{(m+1)!}{2} + m\Gamma'(2m+1) + (2m)(2m)! \ln\rho \biggr] \biggr \} &\nonumber\\
\fl
 - \biggl \{ (\Omega - 1)\left[i(\mathfrak{p}^*\mathfrak{z}_0 -\mathfrak{z}_0^*\mathfrak{p} ) + m + 2|\mathfrak{z}_0|^2\right] \biggr \}, & \label{eq:Efunctional}
\end{eqnarray}
where $\Gamma'(x)$ is the derivative of the gamma function for a positive integer $x$. The interatomic energy term demonstrates the self-bound feature by equilibrating the spatial attraction and repulsion. In the last parenthesis, the wavepacket repulsion dynamics, quantum phase imprinting, and rigid-body revolution are induced in the rotational field.

\subsection{Density Profiles and Phase Portraits}

The ground state properties of the rotating QD were investigated by stirring the condensate adiabatically to ensure equilibrium at a certain $\Omega$ through re-thermalization processes. By minimizing the energy function in Eq.~(\ref{eq:Efunctional}), the solutions of the coupled equations $\partial \mathscr{E}/\partial \rho = 0$ and $\partial \mathscr{E}/\partial R = 0$ yield the optimal topological quantum number $m$, wavepacket width $\rho$, and center-of-mass position $R$ which reveal the effects of confinement, nonlinear interaction, and rotational speed on the ground state configuration of the QDs.

\begin{figure}[t!]
\begin{center}
\includegraphics[width=0.8\textwidth]{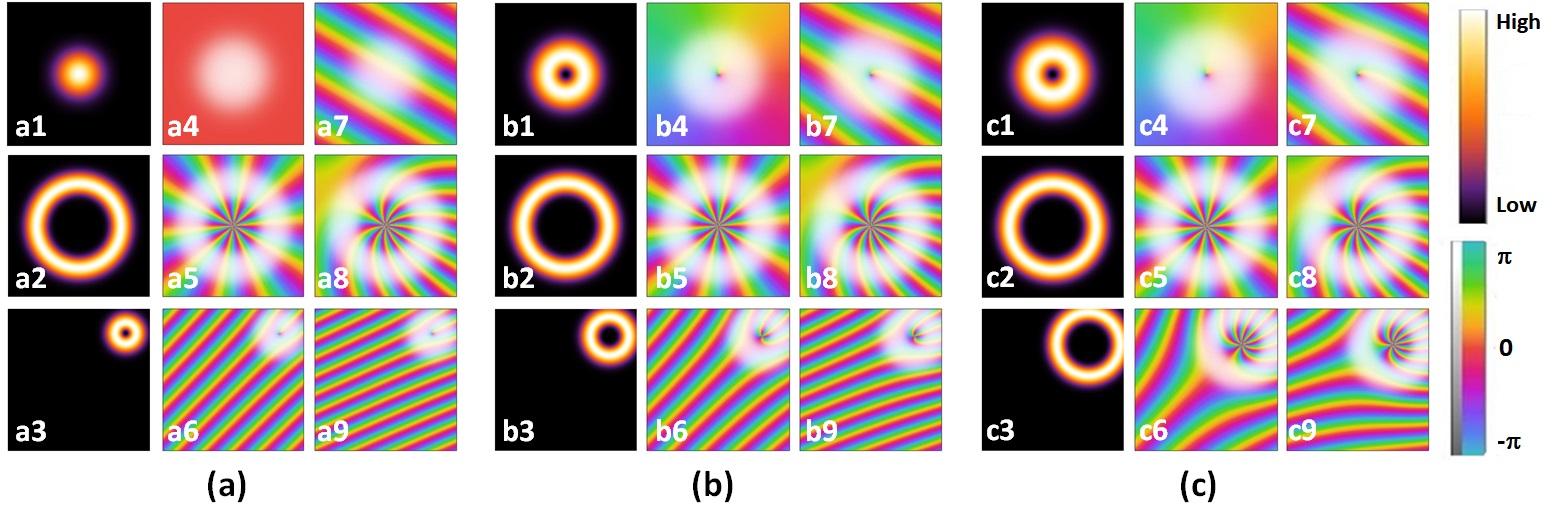}
\caption{(Color online) For (a) $N = 20$, (b) $N = 60$, and (c) $N = 100$, the left columns show density profiles of ground state QDs at $\Omega = 0.6, 1.4, 1.8$ from top to bottom. The corresponding phase portraits of QDs with different linear phase parameters $\vec{p}(0) = 0$ and $\vec{p} = (\sqrt{2}, \sqrt{6})$ are shown in the middle and right columns, respectively.     
}\label{Fig1}
\end{center}
\end{figure}

For QDs with particle numbers $N = 20\,(a), 60\,(b)$, and $100\,(c)$, the ground state density profiles at $\Omega = 0.6, 1.4$, and $1.8$ from top to bottom are shown in the left columns of Fig.~\ref{Fig1}.  
At slow rotations, when the particle number is low, the bead state with $m = 0$ and $R = 0$ is an energetically favorable ground state, as shown in Fig.~1(a1). When the particle number increases, the repulsive LHY interaction plays a decisive role in the formation of vortex states with nonzero $m$, such as the doughnuts shown in Figs.~1(b1) and (c1), which correspond to $m = 1$. For these two states, the respective widths of the wavepacket are $\rho = 1.3$ and $\rho = 1.4$, effectively showing the mixing of higher Landau levels. 
As we increased the rotational speed to $\Omega = 1.4$, Figs.~1(a2), (b2), and (c2) show the tendency of formation of the giant vortices with $m \geq 10$ and large cores owing to the strong centrifugal force. Consequently, along with the generation of large surface tension in the presence of the peculiar nonlinear interaction between liquid atoms, the narrowing of the wavepacket to $\rho < 1$ is evident.
On the other hand, at even faster rotations with $\Omega = 1.8$, the system prefers to form off-center vortex states with nonzero $R$ instead, as shown in Figs.~1(a3), (b3), and (c3). For these deflected QDs, the cores of the vortices increased with increasing particle number, but were lower than those of the vortex states, as mentioned earlier. This phenomenon can be attributed to the counterbalance between the centrifugal force and intense compression of the anharmonic potential. 

In the middle and right columns, the colored lines depict the spatial phase distributions of the QDs subjected to a finite drive. \textcolor{black}{Without loss of generality, the drive is chosen not to point in a typical symmetric direction, for example, $\vec{p}(0) = (\sqrt{2}, \sqrt{6})$.}
When $\vec{p}(0) = 0$, the bead QD was pinned and exhibited a spatially homogeneous phase distribution, as shown in (a4). By contrast, the plane wavefronts displayed in (a7) reveal the tendency of wave propagation driven by an instant linear drive. For the vortex and off-center vortex states, the topological charge number $m$ is the repetition rate of the phase variation in $2\pi$ within the radial-like [middle row: (a5) and (a8), (b5) and (b8), (c5) and (c8)] and fork-like patterns [(b7), and the bottom row: (a6) and (a9), (b6) and (b9), (c6), and (c9)]. Launching an instant impulse produces strain on the surface of the QDs and induces additional phase imprinting upon its wavefunction, which can modify the wavefront slope.

\subsection{Analogy with Optical Vortices}

A brief comparison between rotating QDs and optical vortices is useful for interpreting the underlying physics of phase portraits because the two bosonic systems share several critical features. In addition to the spin angular momentum, light fields having azimuthal phase dependence $\exp(i \ell \phi)$ refer to the beams carrying a quantized OAM $\ell \hbar$ per photon \cite{Allen1992}. This concept has been realized and extensively applied in optical trapping and manipulation, super-spatial resolution microscopy, and material processing \cite{Dholakia2002, Hell2009, Hnatovsky2011}. 
In the paraxial approximation, the solution of the Helmholtz equation of a cylindrical symmetric optical vortex measured at the beam waist $w_s$ is the Laguerre-Gaussian function:   
\begin{equation}
\hspace{-2cm}
LG_{p \ell}(r,\phi) =  \sqrt{\frac{2 p !}{\pi(p+|\ell|) !}} \frac{1}{w_s}\left[\frac{\sqrt{2} r}{w_s}\right]^{|\ell|}  \exp [i \ell \phi] \exp \left[\frac{-r^2}{w^2_s}\right] L_p^{|\ell|}\left(\frac{2 r^2}{w_s^2}\right), 
\label{Eq-LG}
\end{equation}     
where 
the radial index $p$ of the number of nodes and the angular index $\ell$ are equivalent to the parameters of the Landau level index $n_l$ and the degenerate states $m$ within the Landau level in Eq.~(\ref{Eq-LL}). Similarly, by referring to Eq.~(\ref{eq-wf}), the role of the angular index $\ell$ is equivalent to that of quantum number $m$ of the QD. Therefore, the optical vortex can be considered to be the classical correspondence of a rotating QD.
%

%
In an optical system, the number of spiral blades and dark bands in a fork-like interference fringe represents the OAM quantum number of the optical vortex. The same rules can be applied to the atomic system of the QDs. Consequently, the number of color bands split from the singular point of the vortex core displayed in the phase portraits in Fig.~\ref{Fig1} can be claimed to be a visual identification of the OAM carried by the QD.
To the best of our knowledge, no study has been conducted on the phase distribution of rotating QDs. Therefore, the analogy with the optical vortices provided in this study suggests rules for further experimental verification, thus inspiring a challenging interferometer architecture for precisely measuring the singular and coherent properties of rotating QDs with adjustable two-species nonlinear interactions.

\subsection{Phase Diagram}
As vortex states with highly quantized circulation are energetically favorable at rapid rotations, the expectation value of the total OAM $\langle L_z\rangle = m + 2|\mathfrak{z}_0|^2$ will increase sharply and continuously with increasing $\Omega$. 
To maintain the bounded QDs at extremely high rotational speeds, the strong attraction between the squeezed atoms inside the thin rings inevitably increased as the core size rapidly expanded. This large expansion causes the QD to collapse in the presence of three-body collisions.
We suggest setting the maximum topological charge number for vortex states to avoid collapse and numerical divergence.

Applying the LLL approximation, the upper bound of the quantum number $m_{\mathrm{max}} = (\Omega_f - \epsilon/2 - 1)/\lambda - 1$, in which $\epsilon = \omega_0^2 - 1$ can be estimated at the cutoff $\Omega_f$. 
Using the parameters $\Omega_f = 2$, $\omega_0 = 0.5$, and $\lambda = 0.1$, we obtained $m_{\mathrm{max}} = 12$. Because of this restriction, the total OAM between the quantum and classical parts is allocated. Instead of forming a vulnerable giant vortex QD with huge topological charges, the formation of an off-center vortex QD with a deflection $R = 2\sqrt{\Omega - \epsilon/2 - 1 - (m+1)\lambda\rho^2}/\sqrt{\lambda}$ may become more energetically favorable beyond a specific rotation frequency. 
Accordingly, the limiting fraction of the cross-sectional area occupied by the vortex cores estimated in the Thomas-Fermi regime for dilute atomic gases \cite{Fisher2003} can also be applied to QDs.    

\begin{figure}[t!]
\begin{center}
\includegraphics[width=0.7\textwidth]{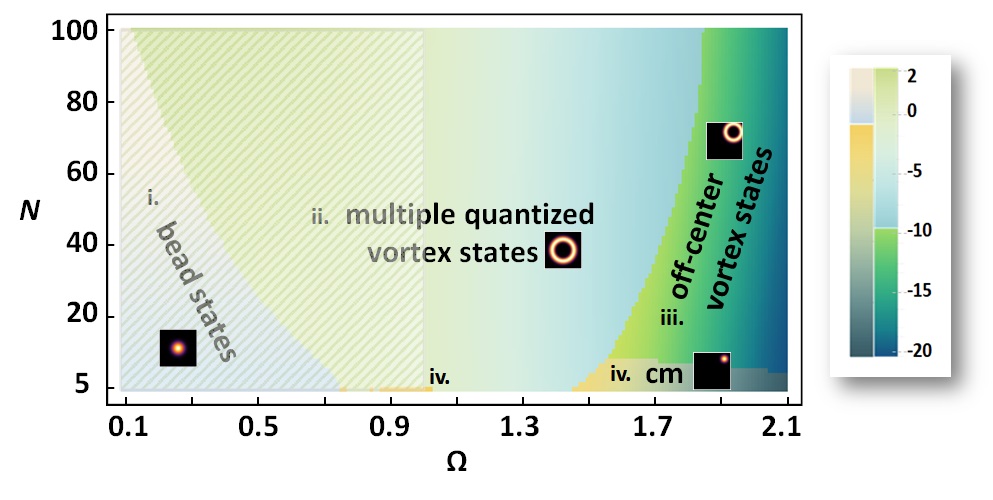}
\caption{(Color online) The $N$-$\Omega$ phase diagram, in which (i) depicts the region corresponding to bead states, (ii) vortex states, (iii) off-center vortex states, and (iv) center-of-mass states. Graphic colors represent the variations in ground state energy.
}\label{Fig2}
\end{center}
\end{figure}

The effects of the nonlinear interaction and rotation speed on the ground state configuration are demonstrated via the $N$-$\Omega$ phase diagram in Fig.~\ref{Fig2}. The graphic colors represent variations in the ground-state energy. Region (i) depicts the bead states with $m = 0$ and $R = 0$, region (ii) depicts the vortex states with $m > 0$ and $R = 0$, region (iii) depicts the off-center vortex states with $m > 0$ and $R \neq 0$, and region (iv) depicts the center-of-mass state with $m = 0$ and $R \neq 0$ embedded in the few-particle and fast-rotation regimes. The color legend provides some references for the ground state energies of the QD with $N$ particles and rotation at $\Omega$. At a fixed rotation speed, the wavepacket width of all quantum states and the energy of the states carrying nonzero topological charges increase with increasing particle number. These growths can be attributed to net repulsive nonlinear interactions. At a fixed particle number, the energy and wavepacket width decrease with an increase in the rotation speed, except for the bead states in region (i), where all parameters are inactive to the environment.  

When the particle number is low, ultra-dilute liquid droplets can be stably sustained if they are energetically more favorable than the atomic cloud subjected to an effective 2D MF interaction given by $V^{2D}(\vec{r}) = \sqrt{8\pi} a_s \delta^{2D}(\vec{r})$. 
Thus, we found that the presence of center-of-mass states approaching $\Omega = 1$ is the signature of dynamic instability in which the gas-liquid phase transition can be induced by particle fluctuations.   

In slow-rotation regime (i), it was also found that the sufficient and necessary condition $({\partial^2 \mathscr{E}}/{\partial\rho^2})({\partial^2 \mathscr{E}}{\partial R^2})-\left({\partial^2 \mathscr{E}}/{\partial\rho \partial R}\right)^2 > 0$ for determining whether a stable QD ground state is violated in the bead states.   
Therefore, the system maintains stability by removing atoms from the dense peak, creating an embedded vortex with multiple circulations during rotation.  
Although it has been reported that singly quantized vortex clusters can be created in QDs at slow rotations \cite{PRL123-160405-2019}, QDs are deformed and probably vulnerable.

\section{Dynamics of Rotating Quantum Droplets}
Depending on the Lagrangian density $\mathcal{L} = i \left( \Psi^* {\partial_t }\Psi - \Psi {\partial_t}\psi^* \right)/2 - \Psi^* H \Psi$, Hamilton's principle of the least action 
\begin{equation}
\delta S[\Psi^*, \Psi] = \int dt \int d\vec{r}\,\, \delta \mathcal{L}(\Psi, \Psi^*,\ldots) = 0, 
\end{equation}
was employed to investigate the QD dynamics and configuration stability. 
%
%
%
%
%
%
%
%
Along with Eqs.~(\ref{eq-wf}) and (\ref{eq:Efunctional}), the Euler-Lagrange equations for each particle and the equations of motion for the characteristic parameters can be derived as follows:   
%
%
%
%
%
%
%
%
\begin{eqnarray}
&&\frac{d}{dt}\left( \frac{\partial \mathscr{L}}{\partial \dot{\mathfrak{p}^*}}\right) - \frac{\partial \mathscr{L}}{\partial \mathfrak{p}^*} = 0 \quad \rm{gives} \quad \dot{\mathfrak{z}_0} = \mathfrak{p} - i \left({\Omega}-1\right). \label{Eq-z0_t}\\
&&\frac{d}{dt}\left( \frac{\partial \mathscr{L}}{\partial \dot{\mathfrak{z}_0^*}}\right) - \frac{\partial \mathscr{L}}{\partial \mathfrak{z}_0^*} = 0 \quad \rm{gives} \quad
\dot{\mathfrak{p}} = -2i \dot{\mathfrak{z}_0} - \left({\omega_0^2}-1 \right)\mathfrak{z}_0 - 2\lambda |\mathfrak{z}_0|^2 \mathfrak{z}_0 \nonumber\\ 
 &&\hspace{2cm} -\left({\Omega}-1\right) \mathfrak{p} + 2\mathfrak{z}_0 \left({\Omega}-1\right) - 2(m+1)\lambda \rho^2. \label{Eq-p_t}
\\
&&\frac{d}{dt}\left( \frac{\partial \mathscr{L}}{\partial \dot{\alpha}}\right) - \frac{\partial \mathscr{L}}{\partial \alpha} = 0 \quad \rm{gives} \quad
\dot{\rho} = \alpha \rho. \label{Eq-rho_t}\\
&&\frac{d}{dt}\left( \frac{\partial \mathscr{L}}{\partial \dot{\rho}}\right) - \frac{\partial \mathscr{L}}{\partial \rho} = 0 \quad \rm{gives} \quad
\dot{\alpha} = -\frac{1}{\mathit{N}(\mathit{m} + 1)\rho}\frac{\partial \mathscr{E}}{\partial \rho}.  \label{Eq-alpha_t}
\end{eqnarray}

\subsection{Periodicity and Stability}

\begin{figure}[hpbt!]
\begin{center}
\includegraphics[width=0.72\textwidth]{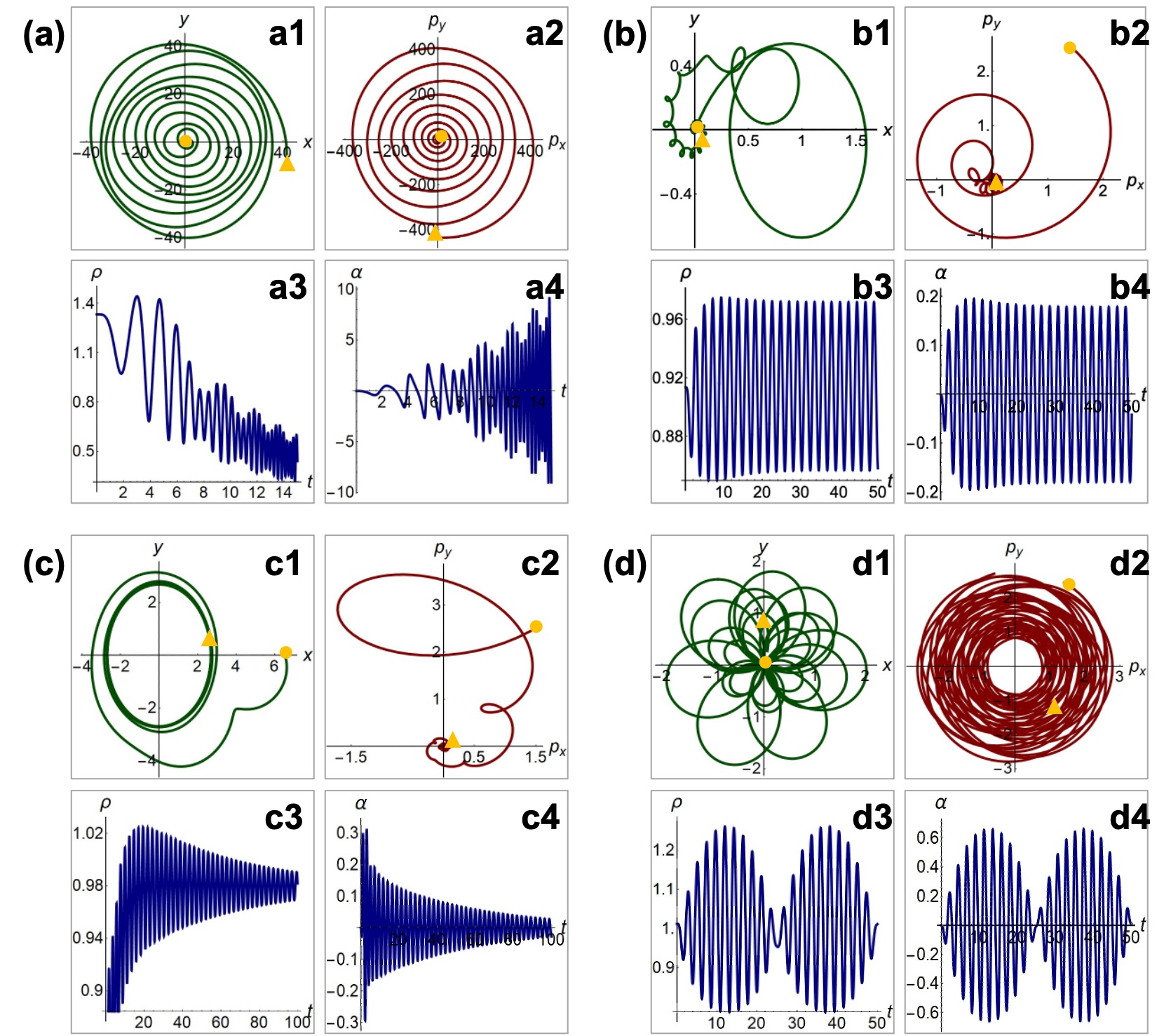}
\caption{(color online) The long-term evolutionary trajectories of the center-of-mass position $\vec{R}$, conjugate momentum $\vec{P}$, variations of width $\rho$, and conjugate curvature coefficient $\alpha$. Under an external drive $\vec{p}(0) = (\sqrt{2}, \sqrt{6})$, (a)-(c) demonstrate the dynamics for $N = 20$ and $\Omega = 0.6$, $1.4$, and $1.8$, and (d) tracks the critical rotation for $N = 100$ and $\Omega = 1.0$. The orange circles on green and brown curves depict the initial $R$ and $P$ of unperturbed QDs, and the orange triangles denote the corresponding parameters of perturbed QDs at some moment after long-term evolution. While bead states are unstable against external perturbations, breathing and self-trapped phenomena can be observed in the vortex and off-center vortex states.}\label{Fig3}
\end{center}
\end{figure}

The coupled equations for $N=20$ and $N=100$, $\Omega = 0.6, 1.0, 1.4$, and $1.8$ were solved in the presence of the initial drive $\vec{p}(0)=(\sqrt{2}, \sqrt{6})$. 
%
\textcolor{black}{In Fig.~\ref{Fig3}, the orange circles on the green and brown curves depict the initial center of mass position and conjugate momentum of the unperturbed QDs, respectively, and the orange triangles denote the corresponding parameters of the perturbed QDs at some moment after the long-term evolution.}

Without the rotational kinetic energy, the bead states were unstable against intense accelerations under the driving field $-(\Omega-1) \mathfrak{p}$ for $\Omega < 1$. As a result, these beads would eventually crash owing to the surge in the center-of-mass position \textcolor{black}{Fig.~\ref{Fig3}(a1), linear momentum 3(a2), curvature 3(a4), and a rapid decrease in the wavepacket width 3(a3).} 

When $\Omega > 1$, the initial drive of the vortex state breaks the axial symmetry, thereby exciting the system to a high-energy state. In contrast to the bead state, the negative contribution $-(\Omega-1) \mathfrak{p}$ plays a crucial role in QD returning to the equilibrium point, as shown in Figs.~\ref{Fig3}(b1) and (b2).
Meanwhile, oscillation phenomena (b3) and (b4) under long-term evolution prove that the ring size is regularly adjusted through the LLM effect to maintain a sufficient surface tension to stably support the vortex state.

Under extremely rapid rotations, Figs.~\ref{Fig3}(c1) and (c2) show that an off-center vortex QD maintains dynamic stability by lowering the topological charge number and shrinking itself (c3) to a solid geometry (c4) to steadily precess around the trap center. In this case, when the steady state is reached after long-term evolution, the self-trapping behavior of the wavepacket to $\rho = 1$ (c3) and the zero variation of the conjugate curvature strongly support the validity of the LLL approximation proposed in this study.

At a critical rotation speed $\Omega = 1.0$, the externally applied effective centripetal force for orbital motion vanishes, leaving a nonzero Coriolis force induced by the velocity variation in the zonal direction, which launches a self-curing rotational motion for the QD. As shown in Fig.~\ref{Fig3}, the quasi-periodic trajectories [(d1) and (d2)] and breathing [(d3) and (d4)] provide evidence of the emergence of collective excitation of the surface mode in the vortex state. 
In the presence of anharmonic trapping and nonlinear effects, the orbit of the QD does not strictly close upon itself after a finite number of oscillations nor does it open. Instead, as the lobes reveal, the QD exhibits multiple-periodic motions between turning points. Similar to that for periodic motion driven by the central-force field, the area enclosed by the moving trajectory is still a rational fraction $2\pi(a_1/a_2)$--after $a_2$ periods, and the radius of the vector of the QD will have made $a_1$ complete revolutions and will have returned to its original position.

Although QDs with multiple quantized circulations are not thermodynamically stable or energetically favorable in homogeneous superfluids or harmonic-confined systems, periodicity and stability analyses have shown that, as the kinetic energy and nonlinear interactions aggregate to comply with the surface tension, vortex configurations in anharmonically-trapped rotating QDs can be stably supported.

\subsection{{Probability Current Density Distribution}}

\begin{figure}[t!]
\begin{center}
\includegraphics[width=0.85\textwidth]{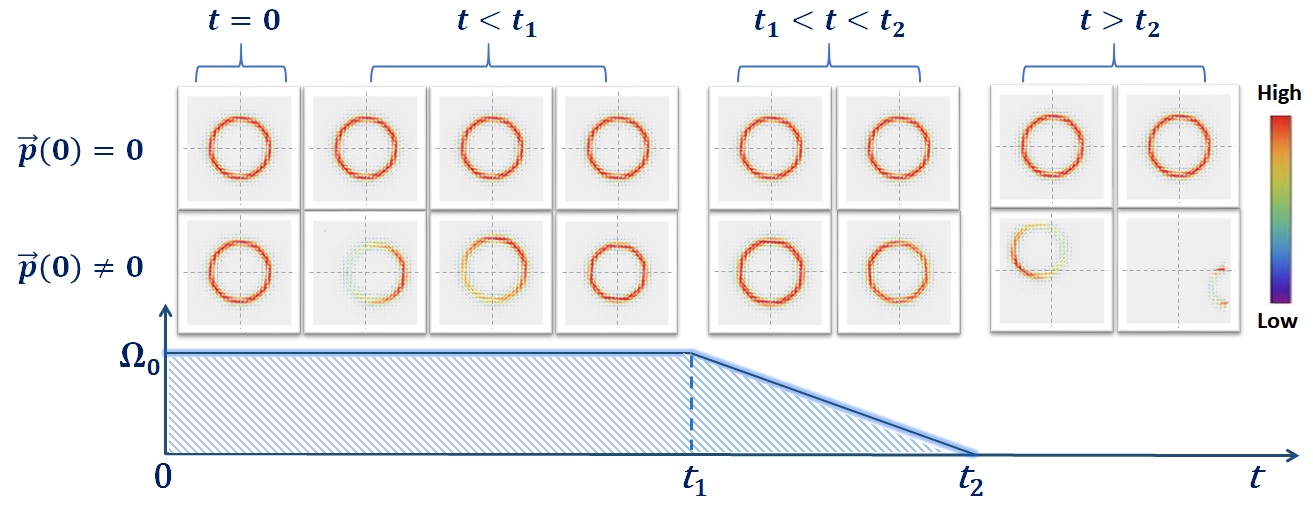}
\caption{(color online) The vector diagrams of the current density distribution for the long-term evolutionary vortex state with initial characteristic parameters obtained at $\Omega_0 = 1.4$. The high-to-low atomic population is marked with rainbow colors ranging from red to blue. The constant rotating field is ramped down in $ t_1 < t < t_2$ and stopped at $t = t_2$. 
{Upper panel:} The superfluids can be sustained in the unperturbed quantum states with phase singularities and quantized topological charges. {Lower panel:} An initial impulse, along with successive expansion and contraction, drives the movement of the vortex but destroys the potential flow within the effective acting time. The destruction persists even after the rotating field is ramped down and stopped.      
}\label{Fig4}
\end{center}
\end{figure}
The rotation at angular velocity $\vec{\Omega}$ can be regarded as a boost to the nonrotating Hamiltonian because $-\vec{\Omega}\cdot \vec{L} = -m \sum_{\bf q} \vec{j}_{\bf q}^\dag\cdot\vec{A}_{\bf q}$, 
where $\vec{j}_{\bf q}^\dag$ is the particle current density fluctuation, and $\vec{A}_{\bf q}$ is the Fourier transform of the transverse artificial vector potential. With the application of the linear response theory, this boost gives rise to a mass current in the condensate frame. For low-energy scattering, the current density of the QD in the presence of a rotating field can be described by    
\begin{eqnarray}
\hspace{-2.5cm}
\vec{j} = \frac{1}{2}\,g_m(\vec{r}, \vec{R})\bigg\{ 4m (\vec{r}-\vec{R})^{-2} \bigg[-(y-y_0)\,\hat{x} + (x-x_0)\,\hat{y}\, \bigg]  
+ \alpha \bigg[ (x-x_0)\,\hat{x} - (y-y_0)\,\hat{y} \bigg]  \nonumber\\
  + \big( p_x \,\hat{x} - p_y \,\hat{y} \big) + \textcolor{black}{(2 \Omega y - y_0)\, \hat{x} + ( x_0 - 2\Omega x)\, \hat{y}}  \bigg\}, \label{eq-current density}  
\end{eqnarray}
where $g_m(\vec{r}, \vec{R}) = 4^{-m} \, |C_m|^2\,(\vec{r}-\vec{R})^{2m} \exp\left[{-(\vec{r}-\vec{R})^2}/2\rho^2\right]$. 
The vector diagrams in Fig.~\ref{Fig4} provide a graphic view of the long-term evolution of the vortex state with initial characteristic parameters obtained at $\Omega_0 = 1.4$. The high-to-low atomic population is marked with rainbow colors ranging from red to blue. The clock flowing vectors characterize the positive sign of the topological charges.  
In the presence of a rotating field, a finite initial impulse $\vec{p}(0)$ along with successive expansion and contraction drive the movement of the vortex within $t < t_1$. 
The variation in the ring reflects quasi-harmonic oscillation under a nonlinear restoring force. Near $t = t_1$, when the vortex center of mass is about to pass through the equilibrium point, a spatially homogeneous distribution of the current density is observed.  

The superfluid behavior was investigated by ramping down the rotating field and ceasing it to zero at $t = t_2$. Surprisingly, when entering the slow-rotation regime with $\Omega < 1$, Eq.~(\ref{Eq-p_t}) shows that any tiny fluctuations induced by the initial impulse can cause considerable phase variation and energy accumulation, which changes the sign of the topological charges, destroy the potential flow, and soon drive the escape of QDs from the trap at $t > t_2$. Accordingly, below the critical rotation speed $\Omega_c = 1$, the vortex states in the shaded area of Fig.~\ref{Fig2} are fragile to external environments and only metastable. As a result, the signature of thermodynamic equilibrium superfluids can only be observed in unperturbed liquid droplets with multiple quantized vortices, as shown in the upper row of Fig.~\ref{Fig4}.

%
In vortex creation experiments for two-component dilute BEC gases \cite{Matthews1999}, where the self-interaction of one component is different from the other, the Hamiltonian is assumed to be invariant under the simultaneous rotation of all hyperfine states. In that work, the observation of annular superfluids was attributed to a lack of common topological stability in a single-component system such as He-4. Concerning the scattering strength imbalance among different species in the effective LHY interaction, this phenomenon can be further studied in asymmetric QD systems by solving coupled LHY-amended GPEs.

\section{Conclusion}

In this study, the stationary properties and dynamics of a rotating QD confined in a two-dimensional symmetric anharmonic trap were investigated. Based on the variational method, we analytically solved the LHY-amended GPE utilizing a trial wavefunction inspired by the work of a 2D Wigner crystal in a strong magnetic field, and addressed the role of the LLM effect. Our study demonstrates the engineering of artificial Lorentz forces in a new type of quantum liquid. 

An embedded vortex with multiple quanta of circulation in the QD is generated by tuning the nonlinear interaction and rotational speed. At even higher rotation speeds, the off-center vortex states were energetically favorable. Some rigid-body-like center-of-mass states can occupy the few-particle regime. By depicting the $N$-$\Omega$ phase diagram, the route for the phase transition was revealed, indicating that manipulating the intensity and phase singularities of rotating QDs and exploring novel singularity-related phenomena in these systems is feasible.

\textcolor{black}{We have provided} a brief comparison of the rotating QD and optical vortex, \textcolor{black}{which is helpful to visualize and interpret} the underlying physics of phase singularities. We consider the number of spiral blades and dark bands in the fork-like interference fringes to be the visual identification of the OAM carried by the QD. To the best of our knowledge, \textcolor{black}{the phase distribution of rotating QDs remains a topic of unexplored research.} Therefore, the rules suggested in this study inspired a challenging interferometer architecture for precise measuring the singular and coherent properties of rotating QDs with adjustable two-species nonlinear interactions.

By applying the Hamiltonian principle and Lagrange dynamics, further investigation of the long-term evolution of rotating QDs allowed us to confirm the stability of quantum states. The self-trapping phenomenon drives the QDs toward the equilibrium point and LLL in rapid rotation regimes, coinciding with the quantum-Hall limit.  
For the special case where $\Omega = 2\omega_0 = 1 $, \textcolor{black}{eliminating the externally applied effective centripetal force for orbital motion gives rise to a nonzero Coriolis force and initiates a self-curing rotational motion of the QD. Appearance of quasi-periodic trajectories and breathing verify the collective excitation of the surface mode in the vortex state.}

The vector diagrams of the current density distributions provide a graphical view of the long-term evolutionary vortex state. \textcolor{black}{External perturbations} induced by the initial impulse can cause \textcolor{black}{great enough} phase variation and energy accumulation \textcolor{black}{to} destroy the potential flow. Therefore, the superfluids can only be steadily sustained in unperturbed quantum states \textcolor{black}{bearing} phase singularities and quantized topological charges.

Sharing many similarities between a quantum Hall system and rotating BECs, applying the same mathematical formalism in this study proved that a QD with multiple topological charges can be stably supported in anharmonic confinement. Our studies have shown the possibility of simulating quantum Hall-type effects in a controlled manner with low-dimensional non-uniform quantum liquids. Systems of rotating BECs with multiple degrees of freedom for tunable parameters are versatile and are promising for topological quantum computations.

\section{Acknowledgement}
We thank the Ministry of Science and Technology, Taiwan, for partial financial support under grants NSTC 110-2221-E-845-004, NSTC 112-2622-E-845-001, and NSTC 112-2112-M-034-001. W. H. K. gratefully acknowledges Prof. Wen-Feng Hsieh for the helpful discussion.


\end{document}